# All-optical helicity-independent switching state diagram in GdFeCo alloys

Jiaqi Wei[1,2,*], Boyu Zhang[1,2,*], Michel Hehn[2], Wei Zhang[2], Gregory Malinowski[2], Yong Xu[1], Weisheng Zhao[1,†], and Stéphane Mangin[2,†]

[1]*Fert Beijing Institute, Beijing Advanced Innovation Center for Big Data and Brain Computing (BDBC), School of Microelectronics, Beihang University, Beijing 100191, China*

[2]*Université de Lorraine, CNRS, IJL, F-54000 Nancy, France*

**Abstract**: Ultra-fast magnetization switching induced by a single femtosecond laser pulse, under no applied magnetic field has attracted a lot of attention during the last 10 years because of its high potential for low energy and ultra-fast memory applications. Single-pulse helicity-independent switching has mostly been demonstrated for Gd based materials. It is now important to optimize the pulse duration and the energy needed to switch a GdFeCo magnet depending on the alloy thickness, concentration. Here we experimentally report state diagrams showing the magnetic state obtained after one single pulse depending on the laser pulse duration and fluence for various GdFeCo thin films with different compositions and thicknesses. We demonstrate that these state diagrams share similar characteristics: the fluence window for switching narrows for longer pulse duration and for the considered pulse duration range the critical fluence for single pulse switching increases linearly as a function of the pulse duration while the critical fluence required for creating a multidomain state remains almost constant. Calculations based on the atomistic spin model qualitatively reproduce the experimental state diagrams and their evolution. By studying the effect of the composition and the thickness on the state diagram, we demonstrated that the best energy efficiency and the longest pulse duration for switching are obtained for concentration around the magnetic compensation.

**Introduction:**

The rapidly increasing social needs for digital information drive the development of magnetic recording technology. In magnetic memory devices, digital information is stored by setting the magnetization of storage medium either "up" or "down". The data writing speed is therefore determined by how fast the setting process could be achived[1-3]. The conventional way to switch the magnetization is by applying a magnetic field, which normally requires high power consumption and complex circuit design[4,5]. Later, various methods for manipulating the magnetization without the magnetic field emerged continuously, including strain[6,7], electric field[8,9], heat[10,11] and polarized current[12,13]. However, the switching process in above cases usually happens via a coherent damping precession of the spins[14,15]. The frequency of the precession is normally in the GHz range and the reorientation process can take several nanoseconds[16,17]. Therefore, a faster writing method is increasingly pursued by researchers.

---

\* These authors contributed equally to this work.

† Corresponding author: weisheng.zhao@buaa.edu.com
   stephane.mangin@univ-lorraine.fr

Concerning the effect of ultra-short light pulse on magnetization, the field opened in 1996 with the discovery of ultrafast demagnetization of a Ni film by a 60 fs optical laser pulse[18]. A decade later, Stanciu et al.[19] demonstrated the possibility of using circular femtosecond laser pulses to induce ultrafast magnetization switching in GdFeCo, namely all-optical switching (AOS), which not only removes the need for magnetic field but also drives writing time towards the picosecond timescale. Soon after, it was discovered that a single linearly polarized pulse is sufficient to switch the GdFeCO, which is called all-optical helicity-independent switching (AO-HIS)[20, 21].

The mechanism behind AO-HIS in GdFeCo alloys has been explained by the presence of two magnetization sublattices with two different relaxation times leading to a transient ferromagnetic-like state[20]. Later, Gorchon et al. studied the role of electron and phonon temperatures in AO-HIS[22]. According to these works, the laser fluence should be high enough to sufficiently heat the free electrons, and initiate the demagnetization and switching process. On the other hand, the laser fluence should also ensure that the phonon temperature $T_{ph}$ remains below $T_c$, since crossing $T_c$ would lead to magnetic disorder and the final magnetization state would then be determined by the cooling conditions. Therefore, AO-HIS should only occur within a narrow range of pulse fluence. In addition, the pulse duration is another important laser parameter. As mentioned above, high $T_e$ is essential for AOS, so it is predicted that the maximum pulse duration ($\tau_{max}$) should be on the timescale of the electron-lattice interaction $\tau_{e-l}$ (around 2 ps in GdFeCo[23]) in order to reach an overheating of the electrons. However, AOS with a pulse duration up to 15 ps was reported [22]. Moreover, two recent works on AO-HIS in GdFeCo have demonstrated that $\tau_{max}$ is significantly influenced by the Gd concentration[23,24].

Despite several works investigating the effect of single-shot AO-HIS, the influence of the laser parameters such as fluence and pulse duration and the properties of the GdFeCo sample such as the concentration and the thickness, has not been systematically investigated. In this work, we report the AO-HIS state diagrams for GdFeCo thin films for different compositions and thicknesses by determining the magnetic configuration as a function of laser pulses duration and fluence. These state diagrams allow us to identify the ideal material and beam parameters to obtain an energy efficient switching. We also define the critical switching fluence ($F_{Switch}^C$) as the minimum fluence allowing the switching of a given alloy thin film. It is usually obtained for the shortest pulse duration. We demonstrated that $F_{Switch}^C$ reaches a minimum around the magnetization compensation point. The state diagram is also significantly influenced by the sample thickness. A thinner film allows to increase the maximum pulse duration necessary to achieve AO-HIS. These findings on the influence of the laser parameters and thin film properties on AO-HIS allow a better understanding of the fundamental mechanism behind this switching.

## Results

### Sample structure and characterization

We prepared a series of 10 nm and 20 nm thick $Gd_x(FeCo)_{100-x}$ layers with different Gd concentrations ranging from 22% to 27%. The amorphous alloys are ferrimagnetic metallic materials, with two antiferromagnetically exchange coupled sublattices. The net magnetization of the alloy is given by the sum of the transition metal (FeCo) sublattice magnetization and the rare-earth (Gd) sublattice magnetization. Therefore, by tuning the composition of the $Gd_x(FeCo)_{100-x}$, it is possible to modulate the magnetic properties. For a composition called the magnetization compensation point ($x=x_{comp}$), the net magnetization reaches zero and the coercivity diverges.

The investigated samples were deposited by DC magnetron sputtering onto a glass substrate according to the following multilayered structure: Glass/Ta (3 nm)/Pt (5 nm)/$Gd_x(FeCo)_{100-x}$ (t nm)/Ta (5 nm). The thin Ta capping layer on top prevents the oxidation of the magnetic film, while it allows probing of the magnetic properties via magneto-optical Kerr effect (MOKE). The bottom Ta layer improves adhesion of the structure to the glass substrate. According to the magnetic hysteresis loops obtained by MOKE (see Supplementary Note 1), all studied samples show strong perpendicular magnetic anisotropy (PMA). From the MOKE results, at room temperature, we could determine $x_{comp}$ between 24 % and 25 %. The net magnetic moment is thus aligned in the direction of the FeCo magnetization sublattice below $x_{comp}$ while it changes its sign and becomes aligned with the Gd magnetization sublattice above $x_{comp}$. Note that $x_{comp}$ also depends on the temperature.

### Magnetization state diagram of GdFeCo

Fig.1 (a) shows magneto-optical images obtained on a 20 nm $Gd_{24}(FeCo)_{76}$ film after exposure to a single laser pulse for different pulse durations (50 fs, 1 ps and 3 ps) and different fluences. The film is initially saturated under an external magnetic field before exposure. We could observe that above a certain fluence defined as $F_{Switch}$, the magnetization of the GdFeCo switches. Moreover, the spot showing AO-HIS expands as the fluence increases and a similar trend is shown for different pulse durations. It is clear that $F_{Switch}$ depends on the pulse duration. For fluences lower than $F_{Switch}$ the laser has no effect on the magnetic configuration (not shown). Above a given fluence $F_{Multi}$, multiple domains start to appear in the middle of the spot. Consequently, AO-HIS is only observed for a fluence value between $F_{Switch}$ and $F_{Multi}$. In addition, as displayed in Fig. 1(b), fully demagnetized patterns are only found when the pulse duration increases to 4 ps, indicating that deterministic all-optical switching could not be achieved above certain pulse duration whatever the laser fluence.

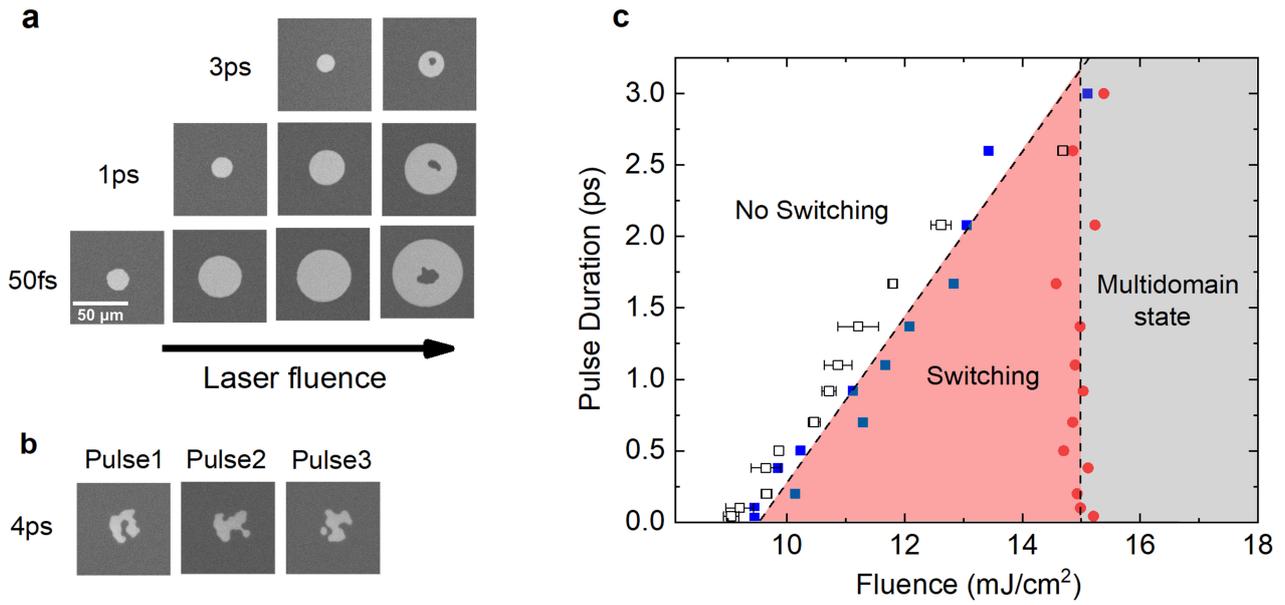

**Fig. 1 Magneto-optical images and all-optical helicity-independent switching state diagram for a 20 nm $Gd_{24}(FeCo)_{76}$ film. a** Magneto-optical images of $Gd_{24}(FeCo)_{76}$ after exposure to a single linearly-polarized laser pulse with a pulse duration of 50 fs, 1 ps and 3 ps, and with various fluences ranging from 9.5 to 15 mJ/cm$^2$. **b** Magneto-optical images of $Gd_{24}(FeCo)_{76}$ after exposure to a single linearly-polarized laser pulse with a pulse duration of 4 ps and a fluence of 17 mJ/cm$^2$. **c** AO-HIS state diagram: Switching fluence $F_{Switch}$ (open black square and full blue square) and multidomain fluence $F_{Multi}$ (full red circle) as a function of the pulse duration for a single linearly-polarized laser pulse. The blue full squares represent the switching fluences $F_{Switch}$ recorded when the diameter of switched area reaches around 10 μm. The open squares are the fitting results obtained via the method proposed by Liu et al.[27] The spatial full-width half-maximum (FWMH) of laser beam is around 70 μm.

Note that the AO-HIS state diagram is very different from that of all-optical helicity-dependent switching (AO-HDS) such as the one obtained by Kichin et al.[25] Indeed, for AO-HIS the fluence window for switching narrows as the pulse duration increases, which is opposite to the trend observed for AO-HDS for which the fluence window broadens as the pulse duration increases. This is one more proof that the mechanism behind the two types of switching (AO-HIS and AO-HDS) are different[26].

To gain a more detailed insight of the dependence of AO-HIS on the pulse characteristics, $F_{Switch}$ and $F_{Multi}$ are presented as a function of the pulse duration indicated by blue squares and red dots, respectively, as shown in Fig. 1(c). The open squares indicate the $F_{Switch}$ obtained via the method proposed by Liu et al.[27] (See Supplementary Note 2). This AO-HIS state diagram allows defining the single pulse laser characteristics leading to AO-HIS, multidomain state or no reversal. Moreover, $F_{Switch}$ and $F_{Multi}$ show obviously different dependence on pulse duration. $F_{Multi}$ is independent on the pulse duration in this pulse duration range, whereas $F_{Switch}$ shows in first approximation a linear increase as the pulse duration increases until $F_{Switch} = F_{Multi}$ which defines the maximum pulse duration ($\tau_{max}$) for which AO-HIS can be observed. Increasing further the pulse duration only lead to a multidomain state.

**AO-HIS state diagrams as a function of the Gd concentration**

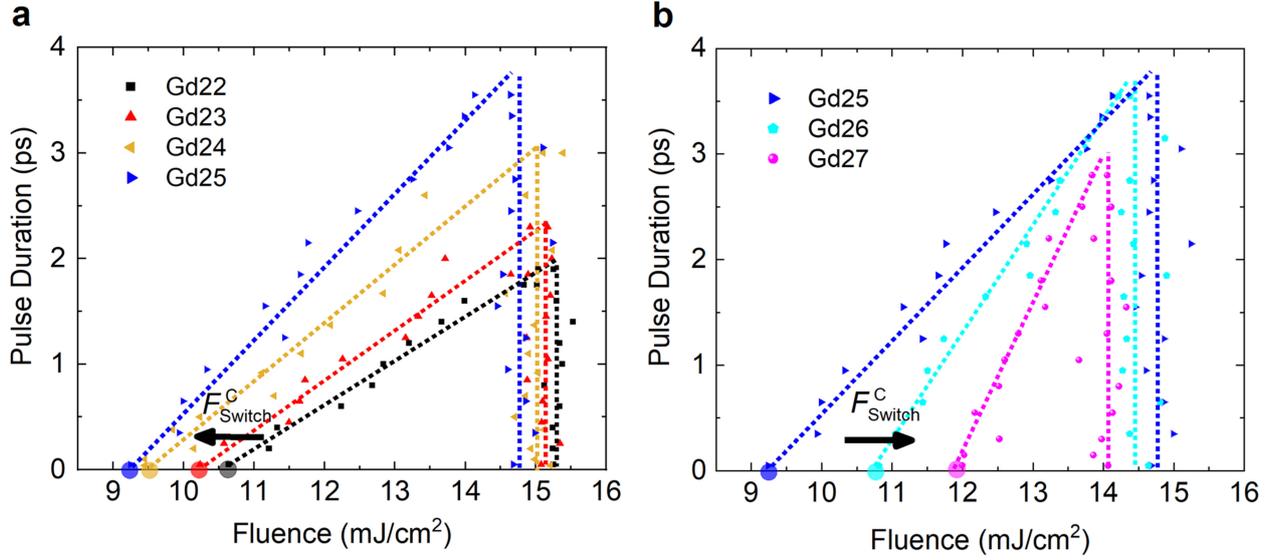

**Fig. 2** AO-HIS state diagrams of 20 nm $Gd_x(FeCo)_{100-x}$ for **a** 22≤x≤25 and **b** 25≤x≤27. The switching regions for different Gd concentrations are highlighted by colored dashed lines. They are mathematically determined by the smallest switching fluence $F_{Switch}^C$ (the x-axis intercept of linear fit of $F_{Switch}$), $F_{Multi}$ and the slope $k$ describing the dependence of $F_{Switch}$ on pulse duration.

Fig. 2 (a) and (b) present AO-HIS state diagrams for 20 nm $Gd_x(FeCo)_{100-x}$ films, with 22≤x≤27 [see Supplementary Figure S7 for state diagrams in 10 nm $Gd_x(FeCo)_{100-x}$]. One can see that the switching regions for different samples share similar outlines (highlighted by colored dashed lines). Indeed, for all concentrations, $F_{Switch}$ increases linearly with the pulse duration and $F_{Multi}$ is independent of the pulse duration. However, obvious changes can be found as the composition is varied. For a more quantitative analysis, the switching region in the state diagram can be characterized by a right triangle. It is mathematically determined by the smallest switching fluence $F_{Switch}^C$ (the x-axis intercept of linear fit of $F_{Switch}$), $F_{Multi}$ and the slope $k$ describing the dependence of $F_{Switch}$ on pulse duration. It is seen that $F_{Switch}^C$ shows a minimum around 25% Gd concentration which is near the magnetization compensation point $x_{comp}$. The slope $k$ increases with the Gd concentration while $F_{Multi}$ decreases which can be attributed to the reduction of the Curie temperature with the Gd composition[24]. We then observe that the fluence range showing AO-HIS is larger and the maximum pulse duration ($\tau_{max}$) is longer around $x_{comp}$. A complete AOS is observed for laser pulse durations up to 3.8 ps for $Gd_{25}(FeCo)_{75}$. This contradicts with the trend predicted by Davis *et al.* that $\tau_{max}$ linearly increases with the Gd concentration[23]. Fig. 3 quantitively shows $F_{Switch}^{C}$ and $k$ as a function of the Gd concentration derived from the AO-HIS state diagrams. It is seen that $F_{Switch}^{C}$ and $k$ share similar trends in 10 nm and 20 nm samples. However, $F_{Switch}^{C}$ scales with the layer thickness as shown in Fig. 3(a); a larger $k$ is obtained for 10 nm GdFeCo at a fixed composition as shown in Fig. 3(b). As a consequence, a complete AO-HIS is observed in 10 nm $Gd_{25}(FeCo)_{75}$ for a pulse duration up to 11.8 ps (see Supplementary Figure S7), which is consistent with the observation of AO-HIS for a 14 nm GdFeCo layer and a pulse duration around 10 ps[22].

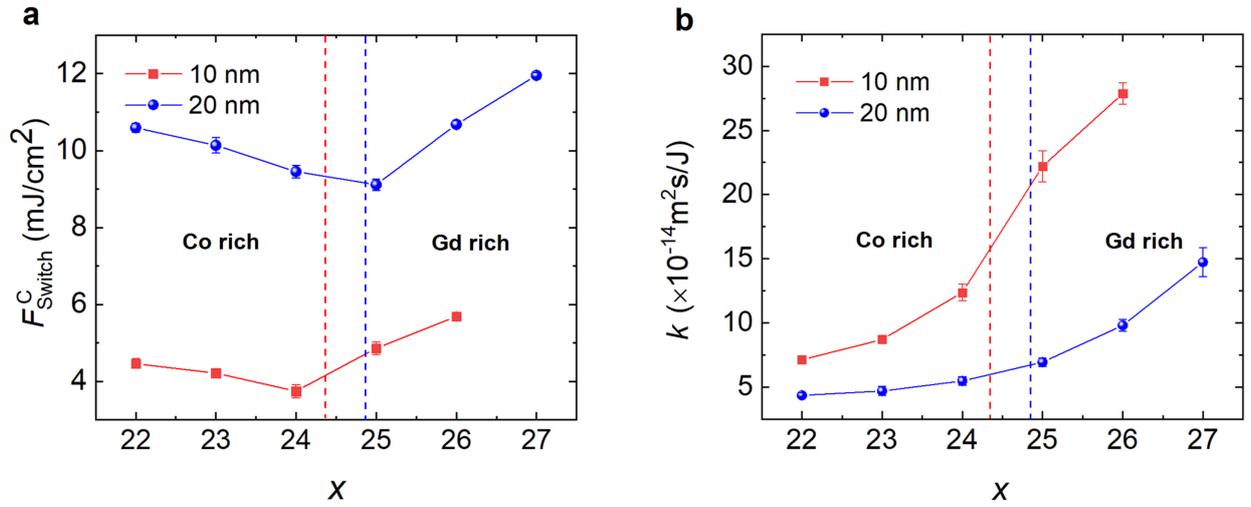

**Fig. 3 Quantitative description of the key features of AO-HIS $Gd_x(FeCo)_{100-x}$ state diagrams. a** The critical switching fluence $F_{Switch}^{C}$ obtained from the intercept of linear fit of $F_{Switch}$ on x-axis is plotted as a function of Gd concentration. **b** Evolution of the slope $k$ extracted from the linear fit of $F_{Switch}$ as a function of Gd concentration.

**Atomistic modeling for single-shot AO-HIS**

In this section, our goal is to test if atomistic modeling can reproduce the experimental results shown above. In the atomistic calculations, each spin is coupled to the temperature of the electron thermal bath. Due to the small heat capacity of electrons, the action of the ultrashort laser pulse first induces a rapid increase of the electron temperature, followed by a slow heat exchange between the electron and phonon thermal baths until equilibrium is reached. The temporal evolution of the electron temperature $T_e$ and the phonon temperature $T_{ph}$ are described by the two-temperature model described in Ref. 28. To simplify calculations, Fe and Co are considered as one sublattice and share the same parameters for atomistic modeling.

We then simulate the temporal evolution of magnetization of Gd and FeCo to an ultrafast light pulse (see Supplementary Note 3). Here, we determine the magnetization state after laser excitation according to the value of z component of FeCo magnetization $m_z^{FeCo}(t)$ when $t$ = 20 ps:1) switching [$m_z^{FeCo}(20) \leq -0.1$], 2) multidomain [$-0.1 < m_z^{FeCo}(20) < 0.1$], 3) no switching [$m_z^{FeCo}(20) \geq 0.1$]. We have chosen "20 ps" to allow for magnetization recovery. Based on such definition, we obtained the simulated state diagram as shown in Fig. 4. It is seen that $F_{Switch}$ greatly increases from 5.16 mJ/cm$^2$ at 50 fs to 7.92 mJ/cm$^2$ at 1 ps while only little change could be observed in $F_{Multi}$, which agrees well with the experimental state diagram. As the pulse broadens, the increase in $F_{Switch}$ is necessary for achieving an overheating of the free electrons which seems to be crucial for AO-HIS. In addition, we calculate the temporal evolution of phonon temperature $T_{ph}$ at $F_{Multi}$ for different pulse durations (see Supplementary Figure S6). Interestingly, the peak $T_{ph}$ shows almost no change for different pulse durations. As a consequence, it depends mostly on the pulse energy and reaches $T_c$ at a very similar fluence for different pulse durations.

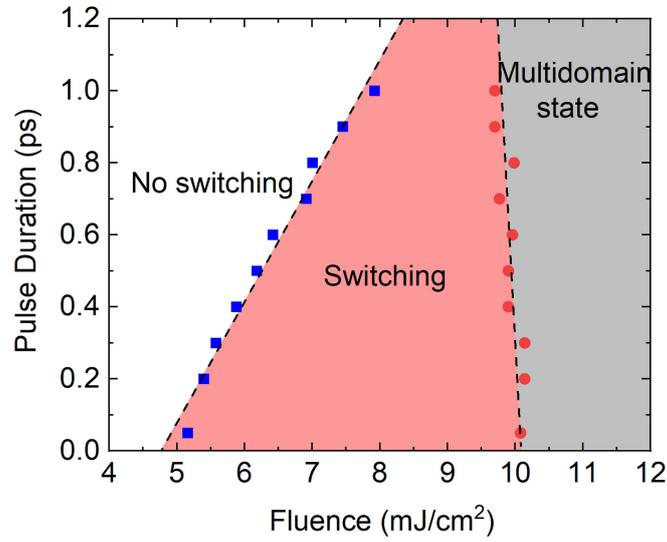

**Fig. 4** Simulated state diagram for Gd$_{26}$(FeCo)$_{74}$. Blue squares and red dots indicate $F_{Switch}$ and $F_{Multi}$, respectively. Dashed lines show linear fits.

Then we try to plot $k$ for different Gd concentrations. As shown in Fig. 5 (a), we calculate $F_{Switch}$ for several pulse durations and compositions. The slope $k$ derived from the linear fit shows a monotonically increase with Gd concentration as presented in Fig. 5 (b). This indicates that when the pulse duration increases by a certain amount, the rise in $F_{Switch}$ required to induce AO-HIS is smaller for the samples with more Gd. This is likely because that the laser heating has a more significant effect with the increased Gd content due to the reduction in Curie temperature $T_c$. Lastly, we calculate $F_{Switch}$ as a function of Gd concentration for 50 fs pulse duration. As displayed in Fig. 5 (c), $F_{Switch}$ depends significantly on the Gd concentration, and reaches a minimum around 32% Gd. This trend is quite similar to the one observed in the experiments. However, the calculated $x_{comp}$ is around 26 (see Supplementary Figure S5) which differs from the composition where the minimum $F_{Switch}$ is obtained. This shift could be explained by a very recent work which shows the position of the $F_{Switch}$ minimum with respect to the Gd-concentration varies with the ratio $\alpha$Fe/$\alpha$Gd utilized in the atomistic modeling[24].

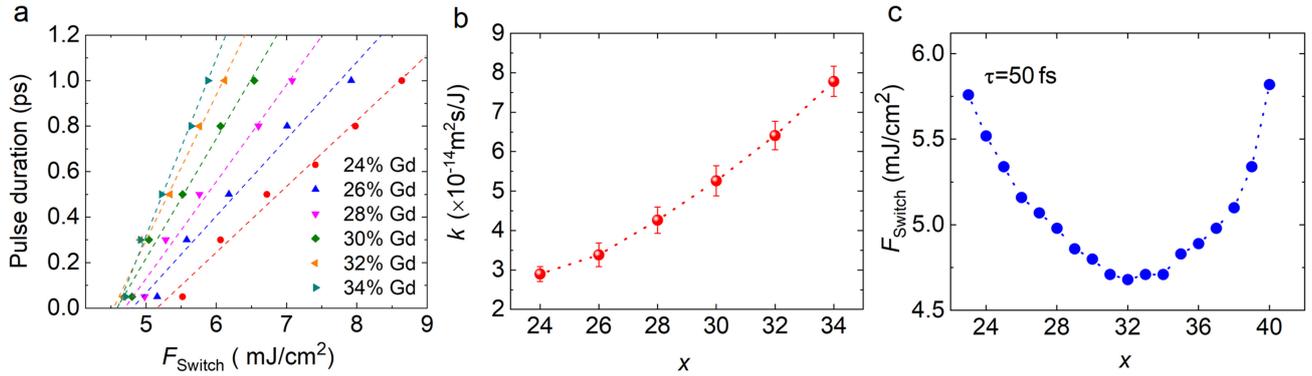

**Fig. 5 Quantitative description of the key features of simulated AO-HIS $Gd_x(FeCo)_{100-x}$ state diagrams. a** Simulated slopes describing the dependence of $F_{Switch}$ on pulse duration for different Gd concentrations. Dashed lines are the linear fits. **b** Slope $k$ as a function of Gd concentration derived from the data shown in (a). **c** $F_{Switch}$ as a function of Gd concentration for a pulse duration of 50 fs.

## Discussion

The experimental AO-HIS state diagrams allow to quickly visualize the laser pulses conditions (Fluence and Pulse duration) required for single-shot AO-HIS. We could then determine the evolution of the state diagrams as a function of the GdFeCo alloy concentration and thickness. This experimental study helps to address technologically relevant controversy and provides crucial guidelines to engineer energy efficient and technologically feasible single pulse all-optical switching of magnetization. Since it has already been shown that both single femto-second light pulse and single femto-second hot electron pulse[29] can induce AO-HIS, the next issue is to generalize the effect to longer pulses to be technologically compatible with feasible electronics. Our study allows answering two controversial questions: How can we optimize the material and the laser excitation to obtain a low energy AO-HIS and how can we observe AO-HIS for long pulse duration?

For the first question, initial studies proposed that the existence of magnetization compensation temperature $T_M$ is essential to achieve AO-HIS in ferrimagnets[30]. Later, it was proved that $T_M$ is not indispensable while low net magnetization $M_{net}$ greatly promotes AO-HIS[31]. However, a following theory work highlighted the importance of temperature derivative of the magnetization $dM_{net}(T)/dT$ and it pointed out that the critical switching fluence is not the minimum at magnetization compensation point $x_{comp}$[32]. In our study, we experimentally demonstrate that $F_{Switch}^C$ significantly depends on the alloy composition and reaches a minimum around $x_{comp}$. We also find that $F_{Switch}^C$ scales with the layer thickness.

The second issue is about the pulse duration threshold $\tau_{max}$. As mentioned above, previous works highlight the importance of dramatic overheating of free electrons. Therefore, it is initially predicted that $\tau_{max}$ should be at the timescale of the electron-lattice interaction $\tau_{e-l}$ (around 2 ps in GdFeCo). However, further studies have shown disparate results ranging from 0.4 to 15 ps[22,23] which could appear surprising at first. In our work, we reveal that $\tau_{max}$ is determined by $F_{Switch}^C$ which defines the base of the triangle-shaped switching region and the slope $k$ which describes the dependence of $F_{Switch}$ on pulse duration and $F_{Multi}$ which seems to be given by the material Curie temperature. To obtain the largest $\tau_{max}$ we should just aim for a small $F_{Switch}^C$, a large $F_{Multi}$ and a large $k$. However $k$ increases with Gd concentration whereas $F_{Multi}$ decreases with Gd concentration and $F_{Switch}^C$ is minimum around the magnetic compensation alloy composition. Therefore, to maximize $\tau_{max}$ it is certainly interesting to be close to the compensation to obtain the smallest $F_{Switch}^C$, but it is also interesting to play with the relative Fe and Co concentration in order to increase the alloys Curie temperature and increase $k$. In addition, a recent work on AO-HIS in GdTbCo highlights that the damping of rare earth site has an obvious influence on $F_{Switch}$[33]. Thus we propose that the magnetization state diagram could also be modified by adding different elements, which will be a subject of future work. The influence of the sample thickness on AO-HIS is more complicated than it appears. From the state diagrams, the right-triangle shaped switching region for 10 nm GdFeCo possesses a much steeper slope than that for 20 nm GdFeCo, which results in a larger $\tau_{max}$. This difference is likely caused by the nonuniform light absorption with respect to the sample depth. Xu *et al.* have calculated the heat absorption profile for similar 20 nm GdFeCo stacks[34]. They revealed that the electron temperature sharply decreases within the bottom 5 nm part of a 20 nm GdFeCo film when excited with a laser having a wavelength of 800 nm. Therefore, it needs much more energy to induce an overheating for thicker samples as pulse duration increases, leading to a smaller $k$.

In conclusion, precise single-shot AO-HIS state diagrams for various $Gd_x(FeCo)_{100-x}$ have been built. It is found that the critical fluences for AO-HIS and multi-domain state exhibit different pulse duration dependences. Calculations based on atomistic spin model are able to reproduce the behaviors observed experimentally. This reveals that electron temperature and lattice temperature play important roles in the all optical magnetization switching. In addition, we demonstrate that the dependence of the critical switching fluence on pulse duration can be tuned by changing the alloy composition and thickness. These findings provide new insights to optimize the materials and laser parameters for energy efficient optical-spintronic devices.

**Methods**

**Atomistic modeling**

In this model, the energy of many-body spin system is described by the Hamiltonian given by[35]

$$\mathcal{H} = -\frac{1}{2}\sum_{i \neq j} J_{ij} S_i \cdot S_j - \sum_i k_u S_z^2$$

where $J_{ij}$ is the exchange integration between spins, $i$ and $j$ ( $i, j$ are lattice sites) giving rise to the magnetic order on the system; $S_i$ is the normalized spin vector and $k_u$ is the local uniaxial anisotropy constant per atom. Meanwhile the Landau–Lifshitz–Gilbert (LLG) equation is used to model the magnetization dynamics of the system:

$$\frac{\partial S_i}{\partial t} = -\frac{\gamma}{(1+\lambda^2)}[S_i \times H_{eff}^i + \lambda S_i \times (S_i \times H_{eff}^i)]$$

where $S_i$ is a unit vector representing the direction of the magnetic spin moment of site $i$, $\gamma$ is the gyromagnetic ratio and $H_{eff}^i$ is the net magnetic field on each spin. It's worth noting that the standard LLG equation is only applicable to simulations at zero temperature. Therefore, Langevin dynamics is introduced here to simulate thermal effects. Then, the net magnetic field in this case is given by

$$H_{eff}^i = -\frac{1}{\mu_s}\frac{\partial \mathcal{H}}{\partial S_i} + \eta_i$$

where $\eta_i$ represents the fluctuations coming from thermal effects and is coupled to $T_e$. Noting that the increase of $T_e$ represents the energy input from laser heating, in order to model the laser heating of the sample, the two-temperature model (2TM) is used:

$$C_e \frac{dT_e(t)}{dt} = -G[T_e(t) - T_{ph}(t)] + P(t),$$

$$C_p \frac{dT_p(t)}{dt} = G[T_e(t) - T_{ph}(t)],$$

$$P(t) = P_0 e^{-(t/\tau_p)^2}.$$

Here $T_e$ is the electronic temperature, $T_{ph}$ is the temperature assigned to the phonon bath, $C_e = \gamma T_e$ and $C_p$ are the heat capacities for electrons and phonons, and $G$ is a coupling parameter between these systems. The parameters we used in the simulation are presented in Table I.

At last, by means of the time integration of the LLG equation through VAMPIRE software package[36], we can obtain the temporal evolution of a system of spins for different pulse energy.

Table I. Parameters for atomistic modeling for GdFeCo. Values are taken from Ref. 37, 38

| | |
|---|---|
| Electronic heat capacity | $C_e = \gamma T_e$, $\gamma = 700$ Jm$^3$K$^{-1}$ |
| Phonon heat capacity | $C_{ph} = 3 \times 10^6$ Jm$^3$ |
| Electron-phonon coupling | $10^{18}$ W/m$^3$K |
| $J_{Gd-Gd}$ | $1.26 \times 10^{-21}$ J |
| $J_{FeCo-FeCo}$ | $4.5 \times 10^{-21}$ J |
| $J_{Gd-FeCo}$ | $-1.09 \times 10^{-21}$ J |
| $\alpha_{Gd}$, $\alpha_{FeCo}$ | 0.01 |

**Author contribution**

W. Z. and S. M. conceived and supervised the project. J. W. and B. Z. initiated the measurements and wrote the manuscript. M. H. developed, grew and optimized the films. W. Z. carried out the atomistic calculation. G. M. and Y. X. helped analyse the data. All authors discussed the results and implications.

**Acknowledgement**

This work was supported by the ANR-15-CE24-0009 UMAMI, by the Institut Carnot ICEEL for the project "Optic-switch" and Matelas, by the Région Grand Est, by the Metropole Grand Nancy, by the impact project LUE-N4S, part of the French PIA project "Lorraine Université d'Excellence", reference ANR-15-IDEX-04-LUE, and by the "FEDER-FSE Lorraine et Massif Vosges 2014-2020", a European Union Program. The authors gratefully acknowledge the National Natural Science Foundation of China (Grants No. 61627813), the Program of Introducing Talents of Discipline to Universities (Grant No. B16001) and the National Key Technology Program of China (Grant No. 2017ZX01032101).